\documentclass[%
 reprint,
 amsmath,amssymb,
 aps,
]{revtex4-2}

\usepackage{graphicx}
\usepackage{dcolumn}
\usepackage{bm}
\usepackage{appendix}

\begin{document}

\preprint{APS/123-QED}

\title{Brillouin lasers in Bragg grating microresonators}

\author{Ryan L. Russell$^{\,1, 2}$}
\email{ryan.russell@sydney.edu.au}
\author{Moritz Merklein$^{\,1, 2}$}
\email{moritz.merklein@sydney.edu.au}
\author{Choon Kong Lai$^{\,1, 2}$}
\author{Cong Tinh Bui$^{\,1, 2}$}
\author{Alvaro~Casas-Bedoya$^{\,1, 2}$}
\author{Duk-Yong~Choi$^{\,3}$}
\author{Stephen~J.~Madden$^{\,3}$}
\author{Benjamin J. Eggleton$^{\,1, 2}$}

\affiliation{$^1$\,Institute of Photonics and Optical Science (IPOS), The University of Sydney, Sydney, NSW 2006, Australia}
\affiliation{$^2$\,Sydney Nanoscience Institute, The University of Sydney, Sydney, NSW 2006, Australia}
\affiliation{$^3$\,Laser Physics Centre, Australian National University, Canberra, ACT, 2601, Australia}

\date{\today}

\begin{abstract}
Chip-scale coherent light sources are required in applications spanning metrology and sensing to telecommunications. Brillouin lasers (BLs) offer a route to ultra-coherent optical sources in compact microresonators with free spectral range (FSR) matched to the Brillouin frequency shift (BFS). However, BFS\,-\,FSR matching typically facilitates cascaded Brillouin scattering, constraining achievable BL output power and coherence. Here, we demonstrate inhibition of cascading in a planar-integrated chalcogenide microresonator by exploiting the photonic bandgap (PBG) associated with a post-fabrication inscribed, reconfigurable intracavity Bragg grating. The PBG inhibits energy transfer within the target Brillouin scattering pathway, such as from pump to first-order Stokes wave. As a quantitative measure of Brillouin scattering inhibition, we report at least six-fold increase in threshold for onset of BL oscillation, which is ultimately limited by thermorefraction. For on-chip pump power of 399\,mW, sufficient for a tenth-order Brillouin cascade, complete inhibition was achieved. Our work positions Bragg grating microresonators as an enabling platform for high performance on-chip BL sources, with reconfigurable modes of operation. 
\end{abstract}

\maketitle

\section{Introduction}
Ultra-coherent electromagnetic sources, particularly in the optical and microwave domains, are cornerstones of modern technologies spanning telecommunications \cite{kikuchi2015fundamentals, pfeifle2014coherent}, sensing \cite{sabri2013toward, ferreira2017roadmap}, microwave photonics \cite{marpaung2019integrated, capmany2012microwave, yao2009microwave}, quantum systems \cite{wehner2018quantum, ludlow2015optical, degen2017quantum, loh2020operation} and ultrasensitive detection, including laser interferometry \cite{collis1970lidar, aasi2015advanced}. 
To meet this demand within a compact footprint, efforts have been focused on chip-scale photonic integration, where stimulated Brillouin scattering (SBS)—a coherent interaction between light and hypersound \cite{brillouin1922diffusion, pant2011chip, merklein2022100}—has emerged as a promising mechanism for optical gain in high-quality (Q) factor resonators \cite{hill1976cw}. Lasing occurs when the Brillouin gain of a Stokes wave exceeds the optical roundtrip loss of the cavity, establishing a well-defined threshold that marks the onset of Brillouin laser (BL) operation. BLs have been demonstrated in a variety of integrated waveguide resonator platforms, including chalcogenide glass \cite{morrison2017compact, kabakova2013narrow, kim2020universal}, which offers exceptionally high SBS gain \cite{eggleton2011chalcogenide, eggleton2019brillouin}, as well as silicon \cite{li2012characterization, otterstrom2018silicon}, lithium niobate \cite{ye2025integrated} and silicon nitride \cite{gundavarapu2019sub}. Sub-Hz-level linewidth BLs can be achieved by exploiting the optomechanical nature of the SBS process \cite{smith1991narrow, gundavarapu2019sub, lee2012chemically}, positioning BLs among the most coherent chip-scale sources. In contrast, the linewidths of monolithically integrated semiconductor lasers are typically limited to kHz level \cite{kim2022consequences}. The optical linewidth narrowing of BL oscillators arises from the combined influence of acoustic damping and optical feedback, which act to filter pump phase noise transfer; this is effective when the photon lifetime exceeds that of the acoustic mode \cite{debut2000linewidth, suh2017phonon}. In the inverse regime, the system operates as a phonon `laser' in which the phonon linewidth is narrowed \cite{grudinin2010phonon, otterstrom2018silicon}. Brillouin oscillators have been deployed in chip-scale laser gyroscopes \cite{li2017microresonator}, optical atomic clocks \cite{loh2020operation} and microwave synthesisers which surpass the performance of state-of-the-art electronic oscillators \cite{li2013microwave, merklein2016widely}.\\

A BL cascade emerges naturally when the Brillouin frequency shift (BFS, $\Omega_B \sim \mathrm{GHz}$) is harmonically matched to the resonator’s free spectral range (FSR), causing each Stokes wave to optically pump the next highest-order mode, which is also supported on a cavity resonance. The seeded Brillouin mode in turn experiences resonant feedback and coherently oscillates as a BL once above threshold. 
In the cascaded regime, the pump power injected beyond the initial threshold is sequentially transferred through the cascade of coupled modes.
Although cascaded BL can be exploited in chip-scale platforms as a source of mutually coherent optical tones \cite{li2017microresonator, li2013microwave, buttner2014phase} and the coveted turnkey Brillouin-Kerr frequency comb \cite{nie2024turnkey, zhang2024strong}, it imposes two fundamental penalties on single-mode BL performance.
First, the pump power is partitioned through the cascade of higher-order BL modes, which restricts scaling of the fundamental BL output power. Second, the many Brillouin scattering pathways introduce additional noise channels; photons in each Brillouin mode can scatter from thermal phonons carrying random phases, effectively coupling each mode to thermal baths which contribute noise \cite{behunin2018fundamental}. Together, power clamping and direct thermally driven noise injection set a lower bound on the achievable linewidth of the fundamental BL--- an effect which is well understood theoretically \cite{suh2017phonon} and verified experimentally \cite{wang2025cascading}.\\

In realising single-mode BL sources that deliver both high output power and ultra-low noise performance, inhibiting cascaded operation remains a challenge. Current mitigation strategies range from lithographic gratings \cite{puckett2019higher, wang2024taming} to tailored cavity mode interactions such as resonance-to-gain detuning \cite{jin2023intrinsic}, a coupled resonator `photonic molecule' \cite{liu2023integrated} or intermodal SBS \cite{wang2021towards, qin2022high}. While effective, the existing methods are either confined to bulk or non-planar architectures, potentially have limited fabrication tolerance, or restrict the BL operating bandwidth. Hence, robust methods for controlling SBS cascading in planar-integrated photonics platforms are sought after. \\

It is well established that scattering and emission processes can be controlled by engineering the photonic environment, as characterised by the optical density of states (DOS) \cite{fujita2005simultaneous, helt2017parasitic, sakoda2005optical}. Through spectral engineering of DOS, the operational state of a BL source can be tailored, such as between single-mode and multi-mode cascaded emission in a single device. A photonic bandgap (PBG) defines a spectral band of vanishing DOS where photon propagation is rigorously forbidden \cite{yablonovitch1993photonic}, analogous to the electronic bandgaps \cite{yablonovitch1987inhibited} responsible for semiconductors. Within the bandgap, the absence of optical modes precludes radiative transitions with final state in the gap and zero-point fluctuations also cease. In our previous work, phase-matched scattering of pump photons into Stokes photons via the SBS process was inhibited in a Bragg grating PBG structure \cite{merklein2015enhancing}. The inhibition effect derives from the `Golden Rule', dictating that transition rates are proportional to the density of available final states \cite{dirac1927quantum, fermi1950nuclear, seitz1950li}; a PBG structure removes these states and so inhibits the transition, as required for energy-momentum conservation. Naturally, this concept can be generalised to control transition processes in other optical systems.\\

In this work, we engineer a Bragg grating PBG structure within a chalcogenide microresonator, offering novel degrees of freedom with which we demonstrate controlled BL operation. The Bragg grating is inscribed all-optically, opening a PBG at specific cavity modes and inducing mode-splitting ranging from 571\,MHz - 1.85\,GHz. We report a high-performance fundamental BL (5.6\,mW threshold, 33\% conversion efficiency) in a resonator of loaded Q-factor $Q_{\mathrm{L}} \sim 7.1\times10^5$, attributed to high SBS gain in our arsenic trisulfide ($\mathrm{As_2S_3}$) waveguides \cite{eggleton2011chalcogenide, eggleton2019brillouin, choudhary2016advanced}. 
We then exploited the vanishing DOS within the induced PBG \cite{sakoda2005optical} to report at least a six-fold increase in the pump threshold for stimulated Brillouin amplification in the cavity. This provides a quantitative measure of BL inhibition arising from inhibited phase-matched Brillouin scattering of the pump into the Stokes mode aligned with the PBG. We observe Stokes amplification to re-emerge at a higher pump power simply because of thermorefractive instability; as the cavity heats under optical loading, all resonance frequencies shift dynamically relative to the pump, causing the Stokes wave to build up in a cavity mode outside the PBG. This can be addressed by active stabilisation of the pump laser to the cavity. However, with a free-running pump we demonstrated the scalability of our PBG-based approach to SBS inhibition in the high-power regime, achieving deep inhibition of a tenth-order Brillouin cascade at 399\,mW of on-chip pump power. 
A notable advantage of photoinduced gratings is their inherent reconfigurability: gratings can be inscribed, erased, and tuned post-fabrication using telecommunication-band light, enabling dynamic control over the microresonator spectral response. This would enable generalised cavity dispersion engineering with applications to parametric oscillators \cite{black2022optical} and frequency comb generation \cite{yu2021spontaneous}. Our work establishes PBG engineering in microresonators as a powerful framework for controlling the operation of BLs with relevance to broader classes of nonlinear oscillators. 

\section{Results}

\begin{figure*}[!t]      
  \centering
  \includegraphics[width=\linewidth]{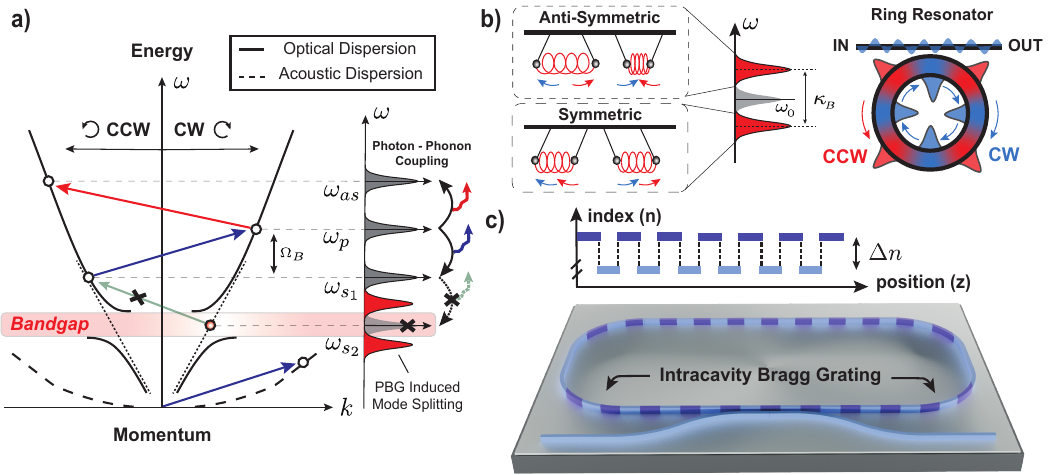}
  \caption{\textbf{Concept of grating-enabled PBG inhibition in microresonator.}  \textbf{(a)} Phase-matching diagram for cascaded Brillouin scattering in resonator. Optical dispersion curves are shown for clockwise (CW) and counterclockwise (CCW) optical (solid) and acoustic (dashed) modes. SBS is mediated by phonons (coloured vectors): CW-oriented pump ($\omega_p$, $k_p$) scatters into CCW-oriented first-order Stokes mode ($\omega_{s_1}$, $k_{s_1}$, blue) which scatters again into CW second-order Stokes mode ($\omega_{s_2}$, $k_{s_2}$, green). Anti-Stokes scattering ($\omega_{as}$, $k_{as}$, red) derives from thermal phonon scattering. Cascaded Brillouin scattering to the mode at frequency $\omega_{s_2}$ is inhibited, where the grating has opened a PBG and induced mode-splitting. It is here that the optical dispersion deviates from a linear `light-line' and group velocity ($v_g$) limits to zero at the band-edge ($v_g = d\omega / d k \rightarrow 0$).
  \textbf{(b)} Illustration of grating-mediated coupling between CW and CCW cavity modes --- analogous to spring-coupled pendulum system. Mode-splitting into symmetric (low frequency) and antisymmetric (high frequency) eigenmodes is shown. \textbf{(c)} Concept of chalcogenide microresonator with intracavity Bragg grating.}
  \label{fig:concept}
\end{figure*}

Backward SBS arises from coupling between two optical modes and a longitudinal acoustic mode, all spatially confined and overlapped within a nonlinear waveguide. As shown in Fig. \ref{fig:concept}(a), SBS between a pump at frequency $\omega_{\mathrm{p}}$ and Stokes at frequency $\omega_{\mathrm{s}}$ is strictly phase-matched, conserving energy and momentum. The pump–to-Stokes frequency shift equals the acoustic frequency, $\omega_{\mathrm{p}} - \omega_{\mathrm{s}} = \Omega_{\mathrm{B}}$, and the wavevector mismatch satisfies $k_{\mathrm{p}} - k_{\mathrm{s}} = k_{\mathrm{a}}$. In our $\mathrm{As_2S_3}$ platform, the BFS of an SBS pump operating at 1550\,nm, determined by the resonantly driven acoustic mode, is approximately 7.7\,GHz. In a cavity, optical modes must satisfy the resonance condition $\phi_{\mathrm{RT}} = n_{\mathrm{eff}}kL = 2\pi m$, where $\phi_{\mathrm{RT}}$ is the round-trip phase, $n_{\mathrm{eff}}$ is the effective refractive index, $k$ the wavenumber, $L$ is the round-trip length, and $m \in \mathbb{Z}$. Realising a BL requires $\Omega_B$ to closely match an integer multiple of the cavity FSR, which is achievable in centimetre-scale resonators \cite{morrison2017compact}. As shown on the right of Fig. \ref{fig:concept}(a), the microresonator cavity can inherently support cascaded SBS because the resonances (grey Lorentzian curves) are spaced periodically at frequencies corresponding to successive Stokes shifts $\omega_n \approx \omega_{\mathrm{p}} - n \, \Omega_B$ ($n \in \mathbb{Z_+}$), enabling phase-matching and gain across multiple orders.

\subsection{Photonic Bandgaps \& Mode Splitting Theory}
\label{sec:mode_splitting}

Key to our approach for controlling BL cascading is the selective inhibition of SBS-mediated energy transfer between resonator modes. We demonstrate this by aligning the target Stokes mode with the narrowband PBG induced by the Bragg grating in the cavity. Within the PBG, the DOS vanishes and Brillouin scattering into the mode is inhibited. It is well understood that reflective structures, such as Bragg gratings, give rise to mode-splitting phenomena in optical resonators \cite{kippenberg2002modal, little1997surface, li2016backscattering, huang2015experimental}. Multiple resonances are typically split if the scattering is incoherent and broadband, such as Rayleigh scattering. We inscribe a coherent Bragg grating reflector that spans the microresonator and is phase-matched to one cavity mode, resulting in singular mode-splitting [Fig. \ref{fig:concept}(a)]. \\

All-optical grating inscription provides a powerful alternative to conventional lithographically patterned gratings \cite{donzella2015design}, leveraging the ability to tune a material’s refractive index through optical exposure and has been demonstrated in chalcogenide waveguides \cite{shokooh2005ultra, lai2023photosensitivity, baker2006sampled, shokooh2006high, saliminia1999photoinduced}.
In $\mathrm{As_2S_3}$, photosensitivity is derived from the near-bandgap absorption of $\sim$ 514\,nm photons \cite{saliminia1999first}, leading to structural rearrangements of the amorphous network and a consequential index change \cite{pfeiffer1991reversible, eggleton2011chalcogenide}. Grating inscription in the telecommunication band ($\sim$ 1550\,nm) has been achieved by addressing the same bandgap through a far weaker three-photon transition \cite{merklein2015enhancing, shen2020reconfigurable, zhu2020photo}. \\

Grating inscription in a microresonator is initiated by exciting a cavity mode in both clockwise (CW) and counter-clockwise (CCW) directions, resulting in a standing-wave intensity pattern along the waveguide. The spatial localisation of high- and low-intensity regions induces a periodic refractive index modulation via the material’s photosensitivity, forming a Bragg grating. To understand how the cavity system responds once the grating perturbation is introduced, we first note that in a typical travelling-wave (TW) resonator such as a ring, the CW and CCW modes are degenerate in frequency. However, once a Bragg grating couples these counter-propagating modes at rate $\kappa_{\mathrm{B}}$, the degeneracy is lifted, and the original mode splits into two standing-wave eigenmodes spectrally split by $\kappa_{\mathrm{B}}$.
These split eigenmodes correspond to symmetric (CW\,+\,CCW) and antisymmetric (CW\,-\,CCW) superpositions of the original travelling waves, forming stationary envelopes with intensity maxima (antinodes) and minima (nodes) that are $\pi$ out-of-phase in space and pinned to the index modulation. Figure \ref{fig:concept}(b) illustrates the analogy that optical mode coupling by a Bragg grating is analogous to two initially degenerate pendulum oscillators coupled by a spring; the coupled system operates in new symmetric and antisymmetric normal modes, with a frequency difference determined by the coupling strength. \\

The self-writing technique used for grating inscription in this work is a by-product of linearity in the Bragg relation \( \lambda_{\mathrm{B}} = 2 n_{\mathrm{eff}} \Lambda \), which expresses that an optical wave couples most strongly to a grating when its effective wavelength in the medium, \(\lambda_{\mathrm{B}} / n_{\mathrm{eff}} \), is twice the grating period $\Lambda$. Thus, an evolving change in $n_{\mathrm{eff}}$ during inscription can be compensated for by the writing wavelength $\lambda_{\mathrm{B}}$ so that $\Lambda$ is maintained. Thus, intracavity Bragg grating formation is inherently self-reinforcing; pumping a split cavity mode excites a standing-wave field that enhances the spatial index modulation and accentuates mode splitting in the course of exposure. Figure \ref{fig:concept}(c) illustrates an intracavity Bragg grating formed in a microresonator. The standing-wave pattern used for inscription produces periodic intensity maxima (antinodes), which induce localised increases in the refractive index (dark blue) along the waveguide. These high-index regions are distinguishable from the lower background index in adjacent regions (light blue). The resulting Bragg grating index modulation profile $n(z)$ is shown. \\

\begin{figure*}[t!]     
  \centering
  \includegraphics[width=\linewidth]{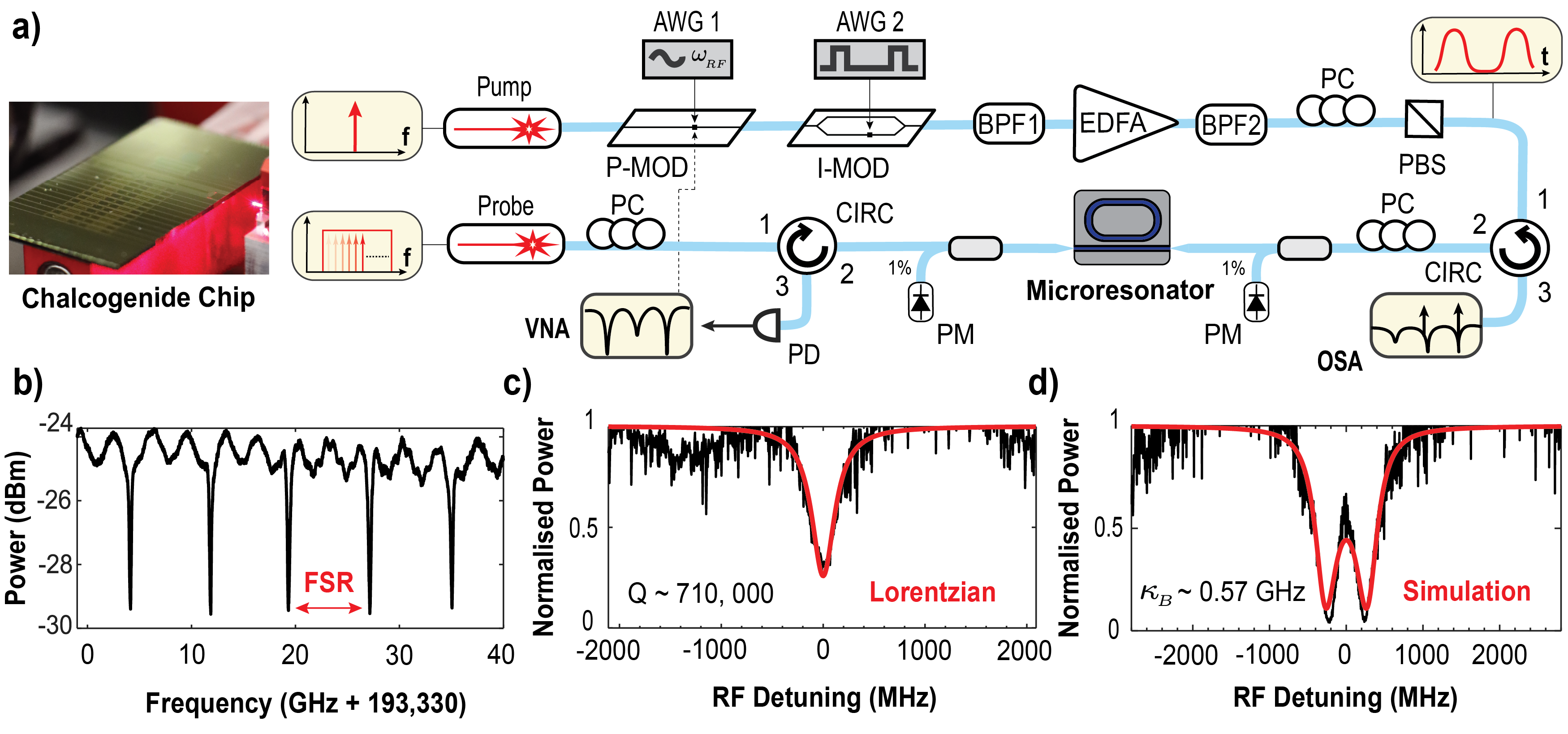}
  \caption{\textbf{Experimental setup, microresonator characterisation and mode-splitting.} \textbf{(a)} Experimental setup and device photo. 
\textbf{(b)} Optical probe transmission spectrum of microresonator with FSR $\sim$ 7.78\,GHz and average extinction ratio $\sim$ 4.9\,dB. 
\textbf{(c)} High resolution normalised transmission response of typical resonance with fitted Lorentzian curve (loaded Q-factor $Q_\mathrm{L} \sim 7.1\times 10^5$). \textbf{(d)} High resolution normalised transmission response of grating-induced resonance splitting with simulated curve [Eq. (\ref{eq:transfer_function})]. P-MOD: phase modulator, I-MOD: intensity modulator, AWG: arbitrary waveform generator, BPF1: narrow ($\sim 20$\,GHz) bandpass filter, BPF2: broad ($\sim 2$\,THz) bandpass filter, EDFA: erbium-doped fiber amplifier, PC: polarisation controller, PBS: polarisation beam splitter, CIRC: circulator, PM: power meter, PD: photodetector, VNA: vector network analyser, OSA: optical spectrum analyser. 
}
  \label{fig:result1}

\end{figure*}
Mathematically, the evolution of clockwise ($A_{\mathrm{CW}}$) and counter-clockwise ($A_{\mathrm{CCW}}$) wave amplitudes in the presence of a grating reflector can be described by coupled mode equations under a slowly-varying envelope approximation \cite{kippenberg2002modal},
\begin{align}
\frac{d A_{\mathrm{CW}}}{dt}
  &= \Bigl(i\,\Delta\omega
     \;-\;\frac{\kappa_{\mathrm{T}}}{2}\Bigr) \,A_{\mathrm{CW}}
     \;+\;i \frac{\kappa_{\mathrm{B}}}{2} \,A_{\mathrm{CCW}} \nonumber\\
  &\quad\quad+\;\sqrt{\kappa_{\mathrm{ext}}}\,s \label{eq:CW_mode_ode}\\[6pt] 
\frac{d A_{\mathrm{CCW}}}{dt}
  &= \Bigl(i\,\Delta\omega
     \;-\;\frac{\kappa_{\mathrm{T}}}{2}\Bigr) \,A_{\mathrm{CCW}}
     \;+\; i \frac{\kappa_{\mathrm{B}}}{2} \,A_{\mathrm{CW}} \label{eq:CCW_mode_ode}
\end{align}
\begin{equation}
    t = \sqrt{\kappa_{\mathrm{ext}}} A_{\mathrm{CW}} - s\quad \quad r = \sqrt{\kappa_{\mathrm{ext}}} A_{\mathrm{CCW}} \label{eq:boundary_cond}
\end{equation}
where $\Delta \omega$ denotes the detuning from the unperturbed cavity resonance angular frequency $\omega_0$, $\kappa_{\mathrm{B}}$ is the Bragg coupling rate, $\kappa_{\mathrm{T}}$ is the total decay rate associated with both the intrinsic loss rate $(\kappa_{\mathrm{int}})$ and external coupling rate $(\kappa_{\mathrm{ext}})$ such that $\tau = 1/\kappa_{\mathrm{T}}$ is the intracavity photon lifetime, $s$ is the input pump field such that $|s|^2$ is the injected power and $t, \, r$ are the transmitted and reflected fields in the bus waveguide, respectively. The photon lifetime constants are related by \( 1/ \tau = 1 / \tau_{\mathrm{int}} + 1 / \tau_{\mathrm{ext}} \). Considering the steady state of the system described by Eq. (\ref{eq:CW_mode_ode} - \ref{eq:boundary_cond}) we obtain the transmission function, 
\begin{equation}
    T(\Delta\omega) = \Bigl|\frac{t}{s}\Bigr|^2 =  \Bigl|  1 - \frac{(\kappa_{\mathrm{T}} / 2 - i \Delta \omega)\kappa_{\mathrm{ext}}}{\kappa_{\mathrm{B}}^2/4 + (\kappa_{\mathrm{T}} / 2 - i \Delta \omega)^2}  \Bigr|^2 \label{eq:transfer_function}
\end{equation}
which has a doublet structure that is prominent in the strong coupling regime ($\kappa_{\mathrm{B}} \gg \kappa_{\mathrm{T}}$). High extinction in transmission is achieved in the bandgap region, centred at $\Delta \omega = 0$. Numerically fitting Eq. (\ref{eq:transfer_function}) to experimental spectra allows the estimation of physical parameters, such as the Bragg coupling rate ($\kappa_{\mathrm{B}}$), intrinsic loss rate $(\kappa_{\mathrm{int}})$, external coupling rate $(\kappa_{\mathrm{ext}})$ and hence intrinsic/loaded Q-factors.

\subsection{Resonator Characterisation \& Mode-Splitting}
\label{sec:BL_inhib}

Figure \ref{fig:result1}(a) shows the experimental setup used for the microresonator characterisation, Bragg grating inscription, and BL experiments. Two counter-propagating optical paths were established for the pump and probe via the circulators on both sides of the chip. A frequency-tunable pump was generated by driving a phase modulator (P-MOD) with a sinusoidal signal at frequency $\omega_{\mathrm{RF}}$ (from AWG1). A quasi-continuous-wave (quasi-CW) pulse train was then generated by driving an intensity modulator (I-MOD) with a square-wave signal (from AWG2). The first-order lower sideband at carrier-offset $\omega_{\mathrm{RF}}$ was isolated (with BPF1), amplified with EDFA, broadband filtered (with BPF2), linearised in polarisation (with PBS) and subsequently aligned to the transverse-electric (TE) axis of the birefringent on-chip waveguide (with PC). High resolution resonator transmission spectra are obtained in the radio-frequency (RF) domain by sweeping the lower sideband offset ($\omega_{\mathrm{RF}}$) with a VNA and then beating the cavity transmission signal with the carrier laser at a PD. In addition, a TE-aligned counter-propagating probe laser was used to analyse the broadband cavity resonance structure. Backward propagating signals relative to the pump wave (odd-order BLs, transmitted probe, reflected pump and even-order BLs) were analysed using an OSA. \\

All experiments in this work were carried out using partially etched chalcogenide microresonator devices. Figure \ref{fig:result1}(b) shows the transmission spectrum of an optical probe swept over a series of five cavity resonances spaced by the FSR $\sim$ 7.78\,GHz. The measured FSR is well-matched to the BFS $\sim$ 7.68\,GHz. The periodic ripple interference appearing near-sinusoidal (amplitude $\sim 1$\,dB) atop cavity resonances (extinction ratio $\sim 5$\,dB), is the resonance structure of a low-finesse Fabry-Perot cavity formed in the bus waveguide owing to Fresnel reflectivity of chip facets. Figure \ref{fig:result1}(c) presents the high resolution characterisation of a single cavity resonance in the RF-domain. A Lorentzian function was fitted using a least-squares routine, yielding a full width at half maximum (FWHM) linewidth of 273\,MHz. This corresponds to a loaded quality factor of $Q_{\mathrm{L}} \sim 7.1 \times 10^5$.\\

We proceeded to induce a PBG in the cavity by inscribing an intracavity Bragg grating, as per the theory described in Section \ref{sec:mode_splitting}. This is achieved using a square-wave pulsed writing beam with a repetition rate of 1\,MHz, 5\% duty cycle, and peak power $\sim$ 100\,mW. Grating inscription using pulses of duration $\tau_{\mathrm{pulse}} \approx 50$\,ns minimises thermorefractive nonlinearities while enhancing the peak exposure intensity ($I$). As is generally characteristic of multiphoton absorption processes \cite{monat2011third}, the underlying three-photon mechanism governing photosensitivity in $\mathrm{As_2S_3}$ \cite{saliminia1999first} leads to a refractive index change that scales cubically with intensity ($\propto I^3$). This makes short, high-intensity pulses particularly effective for efficient grating formation. The intracavity power is also enhanced with respect to the bus waveguide, by a factor that can be estimated as $F_{\mathrm{pow}} \approx Q_{\mathrm{L}} \times \mathrm{FSR} \, / \, (\pi \, f_{\mathrm{0}}) \approx 9.58$\,dB, given loaded Q-factor $Q_{\mathrm{L}} \approx 7.1\times10^5 $, resonance frequency $f_{\mathrm{0}} \approx 193.415$\,THz and cavity $\mathrm{FSR} \approx 7.78$\,GHz. This ensures the Bragg grating was strongly inscribed inside the ring cavity, rather than the bus waveguide [Fig. \ref{fig:concept}(c)]. After thirty minutes of exposure while tracking the writing pump to the upper split-mode, we ceased writing. Figure \ref{fig:result1}(d) shows the resulting mode-splitting in the RF-domain. Due to noise in the measured transmission signal, it was first interpolated using local regression and normalised to unity at the interpolants maximum. Pre-processing afforded more robust least-squares fitting of the theoretical transfer function [Eq. (\ref{eq:transfer_function})] to experimental data, yielding $\kappa_{\mathrm{ext}}/2\pi = 217$ MHz, $\kappa_{\mathrm{int}}/2\pi = 123$\,MHz, $\kappa_{\mathrm{T}}/2\pi = 340$\,MHz and $\kappa_{\mathrm{B}} / 2\pi = 571$\,MHz. Other modes remained unaffected by grating inscription, as characterised by Q-factor and coupling condition. This is an advantage of our approach compared to lithographically defined gratings which may impose scattering loss limited Q-factors \cite{liu2022photonic}.

\begin{figure*}[t!]      
  \centering
  \includegraphics[width=\linewidth]{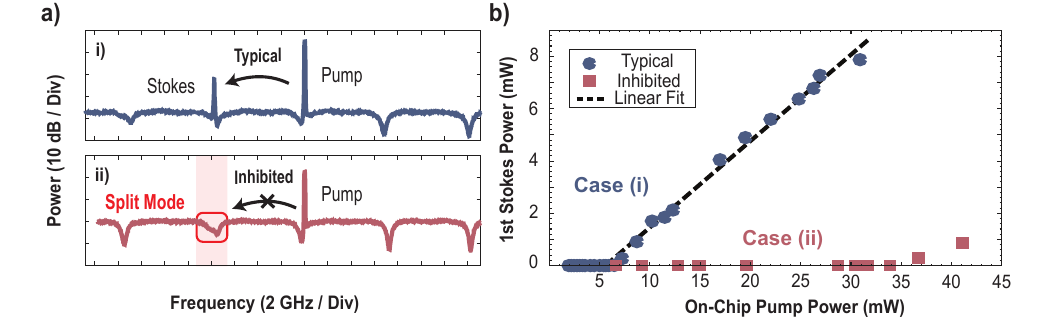}
  \caption{\textbf{Inhibition of Brillouin lasing.} \textbf{(a)} Optical spectra at two pump-resonance detuning states: \textbf{(i)} first-order BL operation between typical resonances with Lorentzian line shape, \textbf{(ii)} BL inhibition with Stokes aligned to the PBG. PBG-induced mode splitting region is shaded red.
\textbf{(b)} First-order BL output power versus on-chip pump power. Blue circles: BL operating as in case (i) - between typical resonances. Red squares: Inhibition of BL gain achieved with Stokes frequency aligned to PBG, as in case (ii).  
}
  \label{fig:result2}
\end{figure*}

\subsection{Brillouin Laser Operation \& Inhibition}
Figure \ref{fig:result2}(a) presents key qualitative observations in two distinct cases (i\,-\,ii) where the pump is detuned across resonance (loaded Q-factor $Q_\mathrm{L} \sim 2 \times 10^5$), with Stokes either aligned to a PBG, or to a typical resonance with Lorentzian line shape. The results were obtained from a microresonator device with FSR $\sim$ 7.35\,GHz, consisting of a 1.9\,um wide, rib waveguide cavity partially etched (33\%) from a 680\,nm $\mathrm{As_2S_3}$ thin film. In case (i), strong Stokes amplification was observed, and the BL operated well above its threshold when driven with peak on-chip pump power of 126\,mW. We then inscribed a Bragg grating and the resulting mode-splitting was given by $\kappa_\mathrm{B} / 2\pi = 1.19$\,GHz. 
At the same pump power, case (ii) confirms the inhibition of BL gain, evidenced by the absence of an observed Stokes signal when aligned in frequency with the PBG. Owing to the still relatively small mode-splitting, a pump detuning of approximately 500\,MHz is sufficient to realign the Stokes wave with the lower-frequency split mode, thereby restoring the Brillouin gain. This detuned regime has implications for the stabilisation of PBG-controlled BL sources, particularly in the presence of thermorefractive effects, which dynamically shift cavity resonances relative to pump and Stokes waves. \\

Having qualitatively observed the BL inhibition effect, we quantified it by measuring the increase in the input power pump required to initiate Stokes wave amplification in the cavity. First, we studied typical BL operation as a baseline for reference. Results were obtained from a microresonator device consisting of a 2\,um wide, rib waveguide cavity partially etched (36\%) from a 1\,um $\mathrm{As_2S_3}$ thin film. A CW pump wave (no pulse modulation) was launched into the fundamental transverse-electric mode of the bus waveguide ($\mathrm{TE}_{00}$) with a lensed-tip fibre. The polarisation of the excited waveguide mode was verified by an infrared (IR) camera and a free-space polarisation beam splitter at the waveguide output. When the pump was aligned in frequency to a typical cavity mode with a Lorentzian line shape (RF-domain spectrum shown in Fig. \ref{fig:result1}(c)), we observed a low-threshold Stokes laser operation at 5.6\,mW of on-chip power, accounting for an insertion loss of $\sim$ 3\,dB at the chip facet. The low threshold is attributed to the high Brillouin gain coefficient of our $\mathrm{As_2S_3}$ waveguides ($\sim 500 \, \mathrm{m^{-1}W^{-1}}$), high loaded Q-factor ($Q_\mathrm{L} \sim 7.1\times10^5$) and relatively small 100 MHz mismatch between cavity FSR $\sim 7.78$ GHz and Brillouin shift $\sim$ 7.68\,GHz, relative to the 273\,MHz FWHM resonance linewidth. A pump-to-Stokes conversion efficiency of 33\% was measured, exceeding that previously reported for comparable chalcogenide microresonators \cite{li2024low, song2024high}. This is attributed to a large ratio between the total optical wave lifetime $\tau_T$ and the external coupling lifetime $\tau_{\mathrm{ext}}$, which governs the fraction of intracavity Stokes power extracted into the output waveguide \cite{loh2015noise}. For on-chip pump power in excess of $\sim 35$\,mW, we observed onset of cascaded SBS. \\

\begin{figure*}[t!]    
  \centering
  \includegraphics[width=\linewidth]{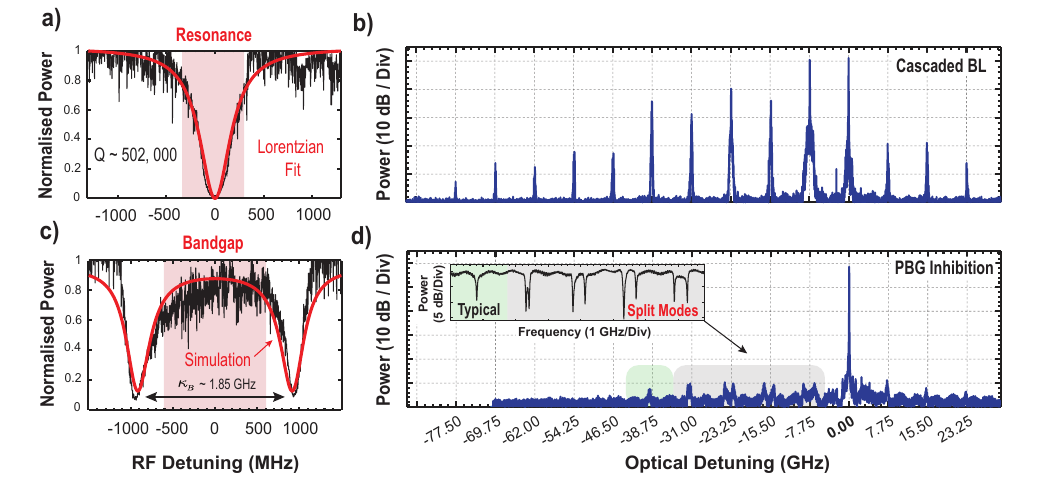}
  \caption{\textbf{High power regime --- PBG inhibition of tenth-order Brillouin cascade.} \textbf{(a)} RF-domain resonance scan fitted to a Lorentzian function.  \textbf{(b)} Optical spectrum showing cascaded Brillouin lasing up to the tenth Stokes order in high-power operation --- detuning is relative to the pump frequency.
  \textbf{(c)} RF-domain resonance scan fitted to split-mode transfer function [Eq. (\ref{eq:transfer_function})]. 
  \textbf{(d)} Optical spectrum showing strong inhibition of the tenth-order Brillouin cascade --- detuning is relative to the pump frequency. Inset: probe optical transmission spectrum showing the mode-splitting of four adjacent resonances.
  }
  \label{fig:result3}
\end{figure*}

We now characterise BL inhibition by aligning the Stokes frequency to the PBG, corresponding to case (ii) in Fig. \ref{fig:result2}(a). As shown in Fig. \ref{fig:result2}(b), no Stokes emission was detected on the OSA (red squares) as the CW on-chip pump power was increased up to 35\,mW, with the Stokes frequency aligned to the split cavity mode identified in the RF-domain spectrum of Fig. \ref{fig:result1}(d). This corresponds to an increase in the BL threshold by at least a factor of six relative to the typical threshold of 5.6\,mW. The observed absence of Stokes amplification is a result of the vanishing DOS at the Stokes frequency within the PBG, which inhibits the phase-matched scattering of pump photons into Stokes photons. Consequently, the optical energy added to the system by increasing the pump power is retained in the pump mode, preventing the build-up of the Stokes field and hence the Brillouin gain. For on-chip pump power in excess of 35\,mW, a weak re-emergence of the Stokes signal was detected. This re-emergence is attributed to thermorefractive shifts of resonances induced by the free-running pump, which causes the Stokes wave to misalign with the PBG. Ultimately, the thermorefractive effects set an upper bound on quantifiable inhibition in this study. With active system stabilisation, such as laser frequency locking, the fundamental-order BL output power can be scaled with a corresponding reduction in the Schawlow–Townes linewidth. This regime is accessed when the second-order BL mode overlaps with the PBG, inhibiting energy transfer away from the fundamental BL mode; this configuration is formally analogous to the pump-to-fundamental BL inhibition presented here.  

\subsection{Brillouin Laser Cascade Inhibition}
We now evaluate the effectiveness of the PBG approach in inhibiting higher-order Brillouin cascading under high pump power conditions. First, to demonstrate the extent of Brillouin cascading currently achievable in our chalcogenide microresonator platform, sufficiently far from the material damage threshold, we operated a pulsed pump at a repetition rate of 1\,MHz, 5\% duty cycle and 399\,mW peak on-chip power, aligned to a typical resonance (loaded Q-factor $Q_{\mathrm{L}} \sim 5.02 \times 10^5$), as shown in Fig. \ref{fig:result3}(a). Under these conditions, we observed a Brillouin cascade extending to the tenth Stokes order on the OSA, as shown in Fig. \ref{fig:result3}(b). Since the pump is quasi-CW, the resulting BL signals are likewise quasi-CW, and the OSA therefore records a time-averaged optical spectrum. Additional signals were observed on the higher-frequency side of the pump, beyond the anti-Stokes line, indicative of weak four-wave mixing (FWM) induced through the Kerr nonlinearity of $\mathrm{As_2S_3}$. We note that the Brillouin nonlinearity dominates Kerr in strength \cite{merklein2015enhancing} and the nonlinear length is short for the normally-dispersive $\mathrm{TE}_{00}$ mode. \\

Since this experiment measures pulse-driven SBS, it is important to clarify the relevant timescales: firstly the pulse duration $\tau_{\mathrm{pulse}} \approx 50\,\mathrm{ns}$ is far longer than both the cavity transit time $t_{\mathrm{transit}} \approx 162\, \mathrm{ps}$ and the mean phonon lifetime $\tau_{\mathrm{phonon}} \approx 15\,\mathrm{ns}$, where $\tau_{\mathrm{phonon}} = 1/\pi \Delta\nu$ \cite{nilsson1971phonon} and $\Delta\nu$ is the Brillouin gain linewidth typically measured with a pump–probe heterodyne method \cite{lai2022hybrid} or inferred directly from the acoustic decay time \cite{merklein2017chip}. Consequently, the system operates in a quasi-CW regime, that is, during the transit of a pulse, the waveguide effectively observes CW operation, with an off-time 20 times longer than the on-time. In the time domain, the BLs inherently also operate in a pulsed regime; however, because the phonon relaxation is fast, it occurs within the pump pulse envelope, and the effective cavity SBS gain remains comparable to that under CW pumping. \\

As the basis for cascaded-BL inhibition, we inscribe multiple spatially coexisting Bragg gratings, each phase-matched to a different cavity mode, so that several adjacent cavity modes are split simultaneously. This structure is formally equivalent to a super-structured Bragg grating (SBG), where the envelope decomposes into orthogonal spatial Fourier components \cite{broderick1997theory, eggleton1994long, eggleton1996nonlinear}. Although a single Bragg grating and its associated PBG would suffice to inhibit BL cascading, the multi-harmonic grating configuration serves to highlight the effectiveness of all-optical inscription for imprinting complex grating structures in microresonators.
A cascaded Brillouin comb generated in the cavity was used as the source for inscription; it was driven by a pulsed pump (1\,MHz repetition rate, 5\% duty cycle and peak on-chip power of 332\,mW). The cavity was exposed in this condition for approximately 20 minutes, while the pump was tracked to the upper split resonance mode. During inscription, a weak thermal self-locking effect was observed \cite{carmon2004dynamical}, providing passive stabilisation and minimising the need for pump frequency adjustments. As shown in the inset of Fig. \ref{fig:result3}(d), the pump and three BL modes independently inscribed grating perturbations, splitting four adjacent cavity modes with a single exposure. The BL mode power decays monotonically with mode-order; we observed that the higher-order BL power was insufficient for grating inscription beyond third order. Figure \ref{fig:result3}(c) shows the RF-domain spectrum of the split mode inscribed by the pump, with simulated curve overlaid (Eq. \ref{eq:transfer_function}). From a numerical least-squares fitting routine, we estimate the physical parameters as $\kappa_{\mathrm{ext}}/2\pi = 213$\,MHz, $\kappa_{\mathrm{int}}/2\pi = 125$\,MHz, $\kappa_{\mathrm{T}}/2\pi = 338$\,MHz and $\kappa_\mathrm{B}/2\pi = 1.85$\,GHz. The larger splitting relative to that shown in Fig. \ref{fig:result1}(d) indicates that a much stronger grating is inscribed here. The grating field reflectivity can be estimated as $r_B = \kappa_{\mathrm{B}} / (2\times \mathrm{FSR}) \approx 0.75$ \cite{shen2020reconfigurable}. Our ability to all-optically inscribe and spectrally tailor complex intracavity Bragg grating structures offers broader utility in applications such as cavity dispersion engineering. \\

Following grating inscription, we restored the pump conditions to those of the baseline tenth-order cascade experiment (1\,MHz repetition rate, 5\% duty cycle and 399\,mW of peak on-chip power). With the pump aligned to a typical cavity mode ($Q \sim 5 \times 10^5$) and Stokes centre-aligned to the split resonance [Fig. \ref{fig:result3}(c)], the OSA spectrum shown in Fig. \ref{fig:result3}(d) indicates no measurable Stokes wave emission. The absence of a detected Stokes signal confirms that the phase-matched Brillouin scattering of pump photons is strongly inhibited by opening a PBG at the first-order Stokes frequency. Consequently, Brillouin scattering into subsequent cavity modes—responsible for initiating cascaded BL—is likewise inhibited owing to the absence of first-order BL. This result demonstrates strong inhibition of Brillouin cascading by the PBG structure in the high pump power regime. The observed non-uniform noise floor likely arises from the weak leakage of the swept probe source through the control attenuator and the result is an artefact of coherent detection in the OSA (APEX OSCA-AP6). Additionally, amplified spontaneous emission (ASE) noise from the EDFA couples into split-mode resonances—superpositions of the CW and CCW travelling-wave cavity modes. The OSA records in the backward direction relative to the pump [Fig. \ref{fig:result1}(a)], hence the ASE may be measurable above the detection noise floor after grating inscription [Fig. \ref{fig:result3}(d)], in contrast to pre-inscription [Fig. \ref{fig:result3}(b)]. Our proof-of-principle demonstration provides the conceptual and experimental basis for inhibiting Brillouin cascading beyond the second order, allowing for high-power scaling of the fundamental-order BL. 

\section{Conclusion}
We demonstrate post-fabrication control of microresonator Brillouin lasers by inscribing an intracavity Bragg grating that opens a narrow photonic bandgap. By aligning the first-order Stokes frequency to the bandgap induced at a cavity mode, Brillouin scattering is inhibited owing to vanishing DOS. This mode-selective PBG inhibition mechanism addresses the challenge of uncontrolled cascaded operation in typical microresonator BL sources, which limits the output power and degrades noise performance \cite{behunin2018fundamental, wang2025cascading}. 
Leveraging the high Brillouin gain of $\mathrm{As_2S_3}$ waveguides and a loaded quality factor of $Q_\mathrm{L} \sim 7.1\times10^{5}$, we realised a microresonator BL with a low threshold (5.6\,mW) and a 33\% conversion efficiency. Aligning the first-order Stokes frequency with the grating-induced PBG raises the lasing threshold by more than a factor of six, unambiguously confirming the inhibitory effect on the Brillouin gain. Quantitative characterisation at higher pump powers is currently constrained by thermorefraction, because the free-running pump is not actively locked to the resonance. Nevertheless, with 399 mW of on-chip pump power—sufficient to drive a tenth-order BL cascade—we still observe inhibition of BL gain. This result highlights the robustness of the PBG inhibition mechanism and the chalcogenide microresonator platform.
Importantly, we show distinct regimes of device operation: Brillouin laser generation typically occurs between 6 - 30\,mW of on-chip pump power, whereas grating inscription requires substantially higher power beyond 100\,mW and at least 20 minutes of exposure. This separation of regimes, in both exposure time and power, combined with the power tolerance and strong nonlinearity of our chalcogenide platform, highlights its suitability for demanding nonlinear photonics applications. Long-term device stability requires that grating inscription does not occur during device operation as a BL source, on any reasonable scale of exposure time or input power. This can be addressed using anti-reflection coatings or heterogeneous integration with our demonstrated $\mathrm{As_2S_3}$\,–\,Germanosilicate inverse taper design, which reduces the facet reflection strength to below –30\,dB \cite{lai2022hybrid}. Although Fresnel reflections are used for mere convenience in this demonstration, counter-propagating writing beams are effective in their absence \cite{ahmad2011photosensitivity}. Consequently, the two regimes of typical device operation and grating inscription become entirely separable. \\

Our demonstration of Bragg grating enabled BL inhibition in a microresonator establishes a path to planar-integrated, single-mode BL sources with high output power and ultra-low noise. Our approach can also be generalised to other classes of nonlinear optical oscillators, offering novel degrees of control. 

\section*{Funding} 
This research was funded by Australian Research Council Grant DP220101431, Australian Research Council Center of Excellence in Optical Microcombs for Breakthrough Science (CE230100006) and US Office of Naval Research Grants (N00014-24-1-2009, N00014-23-1-2597). 

\section*{Acknowledgements}
This research was supported by an Australian Government Research Training Program Scholarship and support from NSW and ACT nodes of the Australian National Fabrication Facility (ANFF), including the Research and Prototype Foundry (RPF) at the University of Sydney. 

\section*{Disclosures}
The authors declare no conflicts of interest. 

\bibliography{main} 

\end{document}